\documentclass[aps,prb,showpacs,superscriptaddress,twocolumn,longbibliography]{revtex4}
\usepackage{amsfonts}
\usepackage{txfonts}
\usepackage{epsfig}
\usepackage{amsbsy} 

\def\be{\begin{equation}} \def\ee{\end{equation}}
\def\bea{\begin{eqnarray}} \def\eea{\end{eqnarray}}

\def\nn{\nonumber}

\def\reff{{\rm eff}}
\def\bk{{\bf k}}

\begin{document}

\title{Topological Hamiltonian as an Exact Tool for Topological Invariants}

\author{Zhong Wang}
\affiliation{
Institute for Advanced Study, Tsinghua University, Beijing 100084, China }

\author{Binghai Yan}

\affiliation{Institute for Inorganic $\&$ Analytical Chemistry,
Johannes Gutenberg University of Mainz,
55099 Mainz,
Germany, EU}


\begin{abstract}
We propose the concept of `topological Hamiltonian' for topological insulators and superconductors in interacting systems. The eigenvalues of topological Hamiltonian are significantly different from the physical energy spectra, but we show that topological Hamiltonian contains the information of gapless surface states, therefore it is an exact tool for topological invariants.
\end{abstract}




\maketitle

\section{Introduction}

Topological insulators are new states of matter with insulating bulk and topologically protected gapless surface, which is robust in the presence of disorder\cite{qi2010a,moore2010,hasan2010,qi2011}. In the noninteracting limit, the existence of gapless surface states can be traced back to the bulk topological invariants \cite{thouless1982,kane2005b,moore2007,Roy2009a,fu2006,fu2007b,fu2007a,qi2008,schnyder2008,ryu2010,wang2010a}, which are defined in terms of the noninteracting Bloch band states. Unfortunately, these topological invariants are inapplicable to systems with electron-electron interaction. For instance, both the definitions of  Thouless-Kohmoto-Nightingale-den Nijs(TKNN) invariant\cite{thouless1982} and the $Z_2$ invariants\cite{kane2005b,fu2006} crucially depend on the concept of ``occupied band'', which is not readily generalizable to interacting insulators.

Recently, topological insulators with strong electron-electron interaction have attracted much interest\cite{raghu2008,shitade2009,zhang2009b,seradjeh2009,pesin2010,fidkowski2010,li2010,dzero2010,rachel2010,wen2010,maciejko2010a,levin2009,zhang2011,hohenadler2011,yu2011,neupert2012,yoshida2012}, therefore, it is highly desirable to find simple and powerful topological invariants for interacting insulators.
For this purpose, Green's function has proved to be a useful tool.  It has been applied to quantum Hall insulators\cite{ishikawa1986,volovik2003}, and has also been proposed \cite{wang2010b} for time reversal invariant  topological insulators in three and two spatial dimensions (3D and 2D). These topological invariants, which are expressed in terms of integrals of Green's function, were identified as the quantized coefficients of topological field theory\cite{qi2008}.
There are numerous recent works making wide use of Green's function\cite{wang2011,wang2011a,gurarie2011,essin2011,yazyev2012}. However, it was a long-standing difficulty that Green's function at the entire frequency domain was needed to obtain topological invariants\cite{ishikawa1986,wang2010b,volovik2003}, which is very laborious in numerical and analytical calculations.

More recently, it has been  explicitly shown\cite{wang2012a} that all topological information of Green's function is encoded in zero frequency, therefore,  Green's function at nonzero frequency is always superfluous for the purpose of topological invariants.
With this observation, a generalized Chern number expressed in terms of zero frequency Green's function was defined for quantum Hall insulators,  which was also generalized to time reversal invariant topological insulators in 3D and 2D\cite{wang2012a}, and also to topological supercondutors\cite{wang2012d}. These topological invariants avoid the difficulty of frequency integral and have found interesting applications in recent works\cite{go2012,wang2012b,budich2012,budich2012a,manmana2012}. In addition to these developments, we would like to mention that a 3D winding number of zero frequency Green's function has been defined in  Ref.\cite{volovik2009,volovik2011} to study the topology of 3D standard model of particle physics.

The applicability of the zero frequency Green's function approach is beyond the ``renormalized bands'' (or quasiparticle) picture.
Generally speaking, the quasiparticle picture is not applicable in the calculation of topological invariants, because the concept of ``quasiparticle'' itself can break down in strongly correlated systems. In the present paper, we will show that the quasiparticle approach to topological invariants can also fail for a different reason. Let us describe this failure in a little more detail.
A qusiparticle with an approximate energy $\omega$ and momentum $\bk$ has a self energy $\Sigma(\omega,\bk)$, therefore, we may expect that $\Sigma(\omega,\bk)$ should play an important role in defining topological invariants. To our surprise, we can see that only $\Sigma(0,\bk)$ appears in the exact topological invariants\cite{wang2012a}. Because electrons are gapped in insulators, it is unexpected that only zero-frequency quantities appear in topological invariants. It is our purpose of the present paper to understand this counterintuitive result.

The central observation of the present paper is that the zero frequency Green's function contains the information of the gapless surface states (e.g. their existence or absence), while the  more conventional ``effective Hamiltonian'', though an accurate tool for bulk energy spectra, fails to do this.  This physical observation confirms the zero frequency Green's function as a crucial tool in calculating interacting topological invariants, which will be of great use in searching for new candidates for topological insulators in systems with strong interaction. As a convenient language,  the concept of ``topological Hamiltonian'' [see Eq.(\ref{topo-H})] is coined for the inverse Green's function at zero frequency , from which various topological invariants can be readily calculated.

The two main conclusions of this paper can be summarized as follows. First, the self-consistent quasiparticle approach is unsuitable for producing topological invariants; second, and more notably, topological Hamiltonian determines the gapless surface states [see Eq.(\ref{main})], therefore it is a general tool for calculating topological invariants.

The rest of this paper is organized as follows. In Sec.\ref{sec:topo-H} we will review the zero frequency Green's function approach, and define the ``topological Hamiltonian''.
In Sec.\ref{sec:eff-H}, we will show that the quasiparticle effective Hamiltonian is unsuitable for calculating topological invariants, then in Sec.\ref{sec:surface} we present the main result of this paper, namely that the counterintuitive topological Hamiltonian can determine topological surface states[see Eq.(\ref{main})]. Lastly, we will make several concluding remarks in Sec.\ref{sec:conclusions}. In the appendix, we list formulas for topological invariants defined in terms of topological Hamiltonian.

\section{Topological Hamiltonian and Effective Hamiltonian}\label{sec:topo-H}

In this section we will briefly review the zero frequency Green's function approach, then we will define the ``topological Hamiltonian'', which is a convenient language for this approach. We then highlight the difference between topological Hamiltonian and the conventional ``effective Hamiltonian''.

Let us fix the notation first.
The many body Hamiltonian is given as $H=H_0 + H_1$, where $H_0=\sum_\bk c_\bk^\dag h_0(\bk)c_\bk = \sum_{\bk,\alpha,\beta}c_{\bk\alpha}^\dag [h_0(\bk)]_{\alpha\beta}c_{\bk\beta}$ is the free quadratic Hamiltonian($\alpha,\beta$ run through all degree of freedom other than momenta, and the chemical potential has been absorbed into the definition of $H_0$) , and $H_1$ term describes the electron-electron interaction. Various (Matsubara, Retarded, etc) Green's functions are defined in the frequency-momentum space in the standard forms\cite{mahan2000}. For instance, the free fermion system has Matsubara Green's function $G(i\omega,\bk)=1/[i\omega-h_0(\bk)]$. For interacting systems, Green's functions are still fully determined by the many body Hamiltonian, though their explicit forms can be very complicated. We note that the Matsubara Green's functions and the retarded Green's functions are equal at zero frequency, therefore either can be used in this paper.

As we have discussed in the introduction, the early topological invariants\cite{ishikawa1986,volovik2003,wang2010b,gurarie2011} constructed from Green's function require integrals over the entire frequency domain, thus they are difficult to calculate. Recently it has been shown mathematically that Green's function at zero frequency contains all the topological information\cite{wang2012a}. Here we would like to briefly review this approach without going into details. The zero frequency Green's function $G(\omega=0,\bk)$ is a Hermitian matrix, therefore, in the eigenvalue equation (at zero frequency) \bea G^{-1}(\omega=0,\bk)|\alpha(\omega=0,\bk)\rangle = \mu_\alpha(\omega=0,\bk) |\alpha(\omega=0,\bk)\rangle \label{mu} \eea all $\mu_\alpha(0,\bk)$ are real numbers.
An eigenvector $|\alpha(0,\bk)\rangle$ with $\mu_\alpha(0,\bk)>0$ is called an ``R-zero''\cite{wang2012a,wang2012}, and the space spanned by R-zeros is called ``R-space''.  We can define the Berry gauge field and Berry curvature in the R-space,  as if the R-space is the ``occupied band(s)'' of a noninteracting insulator. Therefore, we can define topological invariants just in the same way as the noninteracting cases.  For example, for two-dimensional topological insulators with broken time reversal symmetry (namely the quantum Hall insulator),  we have the ``generalized Chern number''\cite{wang2012a}
\bea  C_1 = \frac{1}{2\pi} \int d^2k \mathcal{F}_{xy} \label{chern} \eea where $\mathcal{F}_{ij}=\partial_i \mathcal{A}_j - \partial_j \mathcal{A}_i$, and $\mathcal{A}_i = -i\sum_{\alpha}\langle \bk\alpha |\partial_{k_i}|\bk\alpha\rangle$.  Here $|\bk\alpha\rangle$ runs through an orthonormal basis of the R-space. This is a topological invariant for 2D interacting insulators.
Now we may ask a crucial question: Can we justify discarding Green's function at nonzero frequencies?  Fortunately, the answer is yes. Taking advantage of general mathematical properties of Green's function, it has been established\cite{wang2012a} that nonzero frequencies are always superfluous for the purpose of defining topological invariants, in other words, zero frequency is sufficient.
It is also useful to mention that the physical Hall response  is determined\cite{wang2012a} by this generalized Chern number if there is no nontrivial ground state degeneracy. This approach has been systematically generalized\cite{wang2012a,wang2012d} to topological insulators and superconductors in other symmetry classes.

Since $G^{-1}(0,\bk)$ plays a crucial role in this approach, let us define the `topological Hamiltonian' as \bea h_t(\bk)\equiv -G^{-1}(0,\bk) \label{topo-H} \eea
which can also be written as \bea h_t(\bk) = h_0(\bk)+\Sigma(0,\bk)\eea  following the Dyson equation. In the absence of electron-electron interaction, $h_t$ is reduced to the free Hamiltonian $h_0$. With the definition in Eq.(\ref{topo-H}), the generalized Chern number for interacting insulators, given in Eq.(\ref{chern}), is equivalent to the Chern number of an ``noninteracting system'' with free Hamiltonian $h_t(\bk)$. Therefore, $h_t$ is a convenient language for the zero frequency Green's function approach.
It can be applied\cite{wang2012a,wang2012d} to topological insulators and topological superconductors in all symmetry classes in the periodic table\cite{kitaev2009}. In Appendix \ref{sec:list} we list explicit formulas for these $h_t$ topological invariants to make this paper self-contained.

A naive understanding of $h_t$ is to think it as an effective Hamiltonian for quasiparticles, namely that $h_t$ describes ``renormalized energy bands''.
As we will see, this understanding is incorrect. In fact,
the natural effective Hamiltonian producing the ``(would-be) quasiparticles''\footnote{It is called ``would-be quasiparticle'' because the concept of quasiparticle is precise only when the imaginary part of self-energy is negligible. This condition
is not generally satisfied at nonzero frequency for insulators.} and accurate energy spectra can be described as follows.  Let us start from the self-consistent equations\cite{hybertsen1985,hybertsen1986} for quasiparticle spectra, which can be written in a simplified notation as  \bea [h_0(\bk)+\Sigma(\omega_\alpha,\bk)]|u_\alpha(\bk)\rangle =\omega_\alpha(\bk)|u_\alpha(\bk)\rangle \label{qp}
\eea
which can be more compactly written as \bea G^{-1}(\omega_\alpha,\bk)|u_\alpha(k)\rangle=0 \label{pole} \eea
Eq.(\ref{qp}) is a self-consistent equation with the self energy $\Sigma$ depending on the spectra $\omega_\alpha$.
It is worth to emphasize that it is $\Sigma(\omega_\alpha,\bk)$ instead of $\Sigma(0,\bk)$ that appears in Eq.(\ref{qp}).  In a general interacting system, $\Sigma(0,\bk)$ can be far away from $\Sigma(\omega_\alpha,\bk)$, therefore the topological Hamiltonian approach using $\Sigma(0,\bk)$ is a poor tool for energy spectra. Unlike $\omega_\alpha$ and $|u_\alpha(\bk)\rangle$ given by Eq.(\ref{qp}), which can be interpreted as energy spectra and ``quasiparticles'', the eigenvalues and eigenvectors of $h_t(\bk)$ lack clear physical meaning at this stage.  The powerful aspect of topological Hamiltonian approach is its ability to produce topological invariants rather than energy spectra.

Can we define physically meaningful topological invariants in terms of $|u_\alpha(\bk)\rangle$ (or equivalently, in $\Sigma(\omega_\alpha,\bk)$)? As we will see in the following, this effective Hamiltonian approach turns out to be less fruitful than the topological Hamiltonian approach.
Let us proceed to see how far we can go in the effective Hamiltonian approach. It is natural to define a frequency dependent ``effective Hamiltonian'' \bea h_\reff(\omega,\bk)= h_0(\bk)+\Sigma(\omega,\bk)\eea Furthermore, it seems also natural to define Berry connection in terms of $|u_\alpha(\bk)\rangle$ instead of $|\alpha(0,\bk)\rangle$ given in Eq.(\ref{mu}), since $|u_\alpha(\bk)\rangle$  represents the quasiparticle picture in the interacting system. Now there is a difficulty in doing so, namely that $\Sigma(\omega,\bk)$ is generally not a Hermitian matrix. Let us partially circumvent this difficulty by writing $\Sigma(\omega,\bk)=\Sigma_1(\omega,\bk)+i\Sigma_2(\omega,\bk)$, with both $\Sigma_1$ and $\Sigma_2$ Hermitian. Assuming that the imaginary part $\Sigma_2$ can be safely ignored, we replace $\Sigma(\omega,k)$ by $\Sigma_1(\omega,k)$ in Eq.(\ref{qp}), which now reads  \bea [h_0(\bk)+\Sigma_1(\omega_\alpha,\bk)]|u_\alpha(\bk)\rangle =\omega_\alpha(\bk)|u_\alpha(\bk)\rangle \label{qp-1}
\eea
Here we encounter another difficulty, namely that
$|u_\alpha(k)\rangle$ are generally not orthogonal to each other. This is a relatively minor problem since we can obtain an orthonormal basis  by the Gram-Schmidt process.  For 2D quantum Hall systems, let us define the ``quasiparticle Chern number'' as
\bea  \tilde{C}_1 = \frac{1}{2\pi} \int d^2k \tilde{\mathcal{F}}_{xy} \label{chern-wrong} \eea where $\tilde{\mathcal{F}}_{ij}=\partial_i \tilde{\mathcal{A}}_j - \partial_j \tilde{\mathcal{A}}_i$, and $\tilde{\mathcal{A}}_i = -i\sum_{\alpha}\langle \bar{u}_\alpha(\bk) |\partial_{k_i}| \bar{u}_\alpha(\bk)\rangle$, in which $|\bar{u}_\alpha(\bk)\rangle$ is an orthonormal basis in the space spanned by those $|u_\alpha(\bk)\rangle$ with negative $\omega_\alpha(\bk)$.  The motivation to choose negative $\omega_\alpha(\bk)$ is to generalize the ``occupied band'', which is a crucial concept for the definitions of topological invariants in noninteracting system.

In the remaining parts of this paper, we will show by an explicit example that Eq.(\ref{chern-wrong}) is unsuitable because it can differ from the correct topological invariants given by Eq.(\ref{chern}). We then explain the reason why Eq.(\ref{chern}) instead of Eq.(\ref{chern-wrong}) is the suitable topological invariants from the surface state picture.

\section{Effective Hamiltonian is a poor tool for topological invariant}\label{sec:eff-H}

In this section we study a toy model to show that the self-consistent ``quasiparticle'' approach starting from Eq.(\ref{qp}) is unsuitable for calculating topological invariants. For instance, Eq.(\ref{chern-wrong}) can fail for quantum Hall system.
Let us begin with a two-dimensional (2D) model of quantum anomalous Hall effect\cite{haldane1988,qi2005} with electron-electron interaction.  The free Hamiltonian part $H_0$ describes a two band model\cite{qi2005} with
\bea h_0(k)=\sin k_x \sigma_x +\sin k_y\sigma_y + [m_0 + (2-\cos k_x -\cos k_y)] \sigma_z \label{QAH}\eea
Instead of giving explicit form for the interaction term $H_1$, in the following we will choose an ansatz for self energy $\Sigma(\omega,\bk)$, which is sufficient for our purpose.

Let us make a simple ansatz that $\Sigma(\omega,\bk)\approx\Sigma_z(\omega,\bk)\sigma_z$ near $\bk=0$, in which case the main idea can be most readily appreciated. Near $\bk=0$ we have $h_\reff(\omega,\bk)\approx k_x\sigma_x + k_y\sigma_y + m_\reff (\omega)\sigma_z$, where $m_\reff(\omega)= m_0+\Sigma_z(\omega,\bk=0)$.
Additionally, we assume that the interaction preserves particle-hole symmetry, implying that $\Sigma_z(\omega,0)=\Sigma_z(-\omega,0)$.
From Eq.(\ref{qp})(taking $\bk=0$) we can obtain the self-consistent equation
\bea m_\reff = m_0 + \Sigma_z (m_\reff,\bk=0) \label{eff-mass}\eea This effective mass will be crucial in calculating $\tilde{C}_1$. On the other hand,
the topological mass is defined as
\bea m_t = m_0 + \Sigma_z (0,\bk=0) \label{topo-mass}\eea
Near $\bk=0$ we have $h_t(\bk)\approx k_x\sigma_x +k_y \sigma_y + m_t\sigma_z$.
Following Eq.(\ref{chern}), we can obtain the correct topological invariants as
\bea C_1=\frac{1}{4\pi}\int d^2k \hat{{\bf n}}\cdot(\partial_{k_x} \hat{{\bf n}}\times \partial_{k_y}\hat{{\bf n}}) \eea where the integral range is the Brillouin zone, $\hat{{\bf n}}={\bf n}/|{\bf n}|$, and ${\bf n}=(n_x(\bk),n_y(\bk),n_z(\bk) )$ is extracted from $h_t(\bk)=h_0(\bk)+\Sigma(0,\bk)={\boldsymbol \sigma} \cdot {\bf n}$. As a concrete example, let us pick $m_0=-0.2, \Sigma_z(0,\bk=0)=0.1$, then $m_t=-0.1$, from which it follows that $C_1=-1$. If we pick $\Sigma_z(0.1,\bk=0)=0.3$ (as an example), then $m_\reff=0.1$ is self-consistently obtained from Eq.(\ref{eff-mass}). It follows that $\tilde{C}_1$ determined by Eq.(\ref{chern-wrong}) is $\tilde{C}_1=0$. We knew that Eq.(\ref{chern}) produces the correct topological invariants $C_1$\cite{wang2012a}, therefore, Eq.(\ref{chern-wrong}) fails in this case because $\tilde{C}_1\neq C_1$.

\section{Topological Hamiltonian determines topological surface states}\label{sec:surface}

In the previous sections we have shown that the natural self-consistent $h_{\reff}$ approach starting from Eq.(\ref{qp}) can produce wrong  topological invariants, while the ``topological Hamiltonian'' $h_t (\bk)=-G^{-1}(0,\bk)$ produces the correct topological invariants. This seems mysterious because the eigenvectors $|u_\alpha(\bk)\rangle$ of $h_{\reff}$ [see  Eq.(\ref{qp})] can be interpreted as the ``(would-be) quasiparticles'', while the eigenvectors of $h_t$ have no clear physical meaning.  Why should we expect $h_t$ instead of $h_\reff$ to produce the correct topological invariants?  The answer, as we will show, is that the gapless surface states are determined by $h_t$. The key idea is that the zero energy surface states, if exist, will feel the self energy $\Sigma(\omega=0)$ because they have zero energy. Explicitly, the zero energy surface states $|\psi\rangle$ satisfy the formal equation \bea h_t|\psi\rangle=0 \label{main} \eea which is  self-consistent because it follows from $G^{-1}(\omega)|\psi\rangle= 0$ ( or $[\omega-h_0-\Sigma(\omega)]|\psi\rangle=0$ ) by taking $\omega=0$. Eq.(\ref{main}) remains meaningful when the translational symmetry is broken, because $h_t=-G^{-1}(\omega=0)= h_0+\Sigma(\omega=0)$ can also be defined in coordinate instead of momentum space. Now we can see that $h_t$ is a tailer-made ``Hamiltonian'' for the gapless surface states.
The existence of robustly gapless surface states is a holographical manifestation of the bulk topology, therefore, the direct connection between $h_t$ and surface states firmly establishes the $h_t$ approach to bulk topological invariants.

\begin{figure}
\includegraphics[width=8.0cm, height=7.0cm]{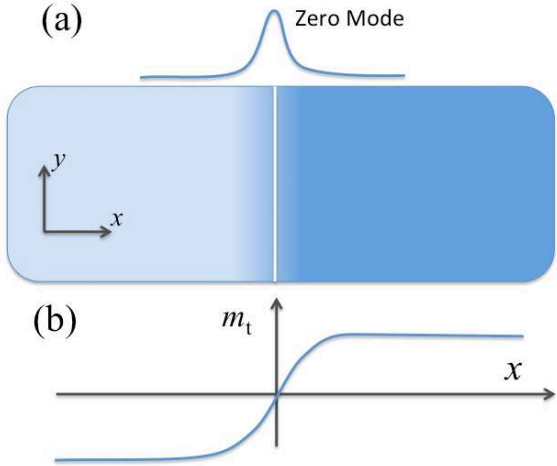}
\caption{The gapless mode on the edge. (a) The physical setting and the zero mode.  (b) $m_t$ as a function of $x$.}
\label{domain}
\end{figure}

Now let us flesh out the above idea in the simple two-band model studied in the previous section.  Consider a 2D system with coordinate $(x,y)$[see Fig.(\ref{domain})]. What we shall show is that at the interface between two bulk states with different $C_1$ (calculated from topological Hamiltonian), there is gapless surface state.  Let us suppose that the many-body Hamiltonian $H$ is a function of $x$, namely that certain parameters (e.g. interaction strength) of $H$ vary as functions of $x$, then the topological Hamiltonian $h_t$ is also a function of $x$. Assuming the form of $h_t$ given in the previous section, we now have $h_t (x)\approx -i\sigma_x\partial_x +\sigma_y k_y + \sigma_z m_t (x)$, since the translational symmetry along $x$-direction is broken. Because it will not affect our conclusion, we take $|m_t(x)|<<1$ to simplify our analysis. Let us suppose that \bea m_t(x\rightarrow +\infty )\rightarrow M \,\,; \,\,\, m_t(x\rightarrow -\infty)\rightarrow -M  \label{M} \eea and that $m_t$ smoothly interpolates between these two limits near $x=0$ [see Fig.(\ref{domain}b)].

Due to the translational symmetry in $y$ direction, $k_y$ is a good quantum number, thus we can study the $k_y=0$ branch first. Now $h_t$ reads

\bea h_t \approx -i\sigma_x\partial_x + \sigma_z m_t(x) \label{hx} \eea
We have omitted terms in higher order of $\bk$ since we are concerned with low energy modes. If $M>0$ [$M$ is defined in Eq.(\ref{M})],
the solution of zero mode from Eq.(\ref{main}) is readily obtained as \bea |\psi(x)\rangle_{\rm zero\,mode}= \frac{1}{\sqrt{2}} \left(
                \begin{array}{c}
                  1  \\
                   i \\
                \end{array}
 \right)  \exp[-\int^x dx' m_t(x')] \label{solution} \eea
The assumption of existence of this zero mode requires that $m_t(x)<0$ when $x\rightarrow -\infty$ and $m_t(x)>0$ when $x\rightarrow\infty$ (otherwise the wavefunction diverges)\cite{jackiw1976,goldstone1981}, which is satisfied by this solution. If $M<0$, then there is also a zero mode solution with $(1,-i)^T$ replacing the $(1,i)^T$ factor in Eq.(\ref{solution}). Note that the existence of zero mode is fully determined by the asymptotic behaviors of $h_t(x)$ as $|x|>>0$, therefore it is determined by the bulk topology. From the above explicit calculation at the interface we have seen that when $m_t(x)$ changes sign at the interface near $x=0$, there are exact zero modes. The above calculation looks in parallel with the noninteracting cases in which we start from $h_0$ instead of $h_t$, however, the crucial difference is that in interacting cases we have a frequency-dependent self energy, and $\Sigma(\omega=0)$ appears in a self-consistent manner.

From the bulk calculation given in the previous section, we have $C_1(m_t>0)=0$ and $C_1(m_t<0)=-1$ (remember that we take $|m_t|<<1$), thus the topological invariants at the two sides of interface are different.  To summarize the above calculations, we have shown that the gapless surface states are associated with the difference of topological invariants ( calculated from $h_t$ ) at the two sides of the interface. In experiment, one side of the interface is often the vacuum, which can be regarded as a topologically trivial insulator, therefore, having ``different topological invariants'' at the two sides of interface means that the insulator has nontrivial topological invariant ($C_1\neq 0$ in this quantum Hall example).

The fact that $h_t$ self-consistently determines the basic properties of gapless surface mode (e.g. the existence or absence of these gapless modes), as given by Eq.(\ref{main}), is among the central results of this paper.  It is equally notable that $h_\reff$ plays no role in the solution of zero energy surface mode. This is the simple reason why $h_t$ instead of $h_\reff$ should determine the bulk topological invariants.

We would like to add a remark for symmetry protected topological insulators, for which the topology is well defined provided that the bulk has the given symmetry. Let us take the time reversal invariant $Z_2$ topological insulators in three dimensions\cite{moore2010,hasan2010,qi2011} as an example. Suppose that the bulk time reversal symmetry is preserved. If the time reversal symmetry at the surface is also preserved, then the relation between $h_t$ and the gapless surface states discussed in this paper is applicable. On the other hand, if the surface states break the time reversal symmetry either by external perturbation(e.g. surface magnetism) or spontaneous symmetry breaking, the surface become gapped, and $h_t$ calculates the coefficient of topological field theory\cite{qi2008}. The same relation between $h_t$ and topological field theory\cite{wang2011b,ryu2012,nomura2012,stone2012} is true for symmetry protected topological superconductors. Therefore, $h_t$ can tell us about the physical topological responses such as the half quantum Hall effect\cite{fu2007a,qi2008} and thermal Hall effect\cite{wang2011b,ryu2012} at the surface of topological states, if the surfaces are gapped.

We also note that we can have another understanding of $h_t$,  if we treat $h_t(\lambda)$ as a function of a tuning parameter $\lambda$ instead of the spatial coordinate $x$. This is an understanding from the perspective of topological phase transition, in which the low energy modes near the transition point play the role of ``gapless surface/interface state'' discussed in this section.

\section{Conclusions}\label{sec:conclusions}

In conclusion, we have shown that for interacting systems, the self-consistent quasiparticle approach given by Eq.(\ref{qp}), which is a natural generalization of the noninteracting ``occupied band'', is unsuitable for the purpose of topological invariants. We have also confirmed that the topological Hamiltonian $h_t$ is the suitable tool, though its eigenvalues are generally very different from the true energy spectra. The simple reason, as we have shown, is that $h_t$ directly tells us about the information of gapless surface states.

Early topological invariants\cite{ishikawa1986,volovik2003,wang2010b} derived from Green's function require integrals over the entire frequency domain, which are difficult to implement. Recently it has been shown\cite{wang2012a} mathematically that only zero frequency is relevant for topological invariants. This result justifies discarding Green's function at nonzero frequencies,  however, a more physical understanding of this approach was absent. We hope that the present paper has filled this gap by looking at the surface states. It is worth noting that the purpose of this paper is to understand the topological Hamiltonian approach as an exact tool, while more realistic models in which this approach may find wide applications remain to be studied.

The $h_t$ approach can be applied\cite{wang2012d} to topological insulators and superconductors in all symmetry classes in the ``periodic table''\cite{kitaev2009}. After obtaining $h_t (\bk)$ for an interacting system, the task of calculating topological invariants is equivalent to doing this for a ``free fermion system'' with Hamiltonian $h_t (\bk)$, thus all noninteracting topological invariants are generalized to interacting systems (See the Appendix \ref{sec:list}).

Let us conclude with the remark that in band structure calculations we are usually more concerned with energy spectra, namely that various versions of the ``effective Hamiltonian'' approach are preferred, however, as we have shown in this paper, such calculation is unsuitable for the purpose of obtaining topological invariants. The topological Hamiltonian approach faithfully produces correct topological invariants, therefore, it is a crucial tool for searching topological insulators/supercondutors with strong electron-electron interaction.

\section{Acknowledgement}

ZW is especially grateful to Shou-Cheng Zhang for fruitful discussions and  collaborations in previous work. ZW also thanks Zheng-Yu Weng for helpful discussions.
ZW is supported by Tsinghua University Initiative Scientific Research Program(No. 20121087986). BY is supported by the ERC Advanced Grant (291472).

\appendix

\section{Topological invariants for interacting (integer) topological phases in general spatial dimensions}\label{sec:list}

In this appendix we list topological invariants for interacting integer topological insulators and superconductors in the periodic table\cite{kitaev2009}. These formulas take the same mathematical forms as their noninteracting counterparts, except that they are expressed in term of zero frequency Green's function [$h_t(\bk)=-G^{-1}(0,\bk)$] for interacting systems instead of the free Hamiltonian $h_0(\bk)$ for noninteracting systems. We list these formulas to make this paper self-contained. The noninteracting counterparts of the listed formulas can be found in previous literatures, e.g. in Ref.\cite{ryu2010}.

First we list topological invariants for the two complex classes,  class A and AIII, which are characterized by integer topological invariants.

For topological insulators in class A in $2n$  dimensions ($n$ is an arbitrary positive integer), we can define the Berry gauge field $\mathcal{A}$ and the Berry curvature $\mathcal{F}$ in the R-space [see Section \ref{sec:topo-H}], which can be intuitively understood as the ``occupied band'' of $h_t(\bk)$, then we can define
the  $n$-th  generalized Chern number as \bea  {\rm A}: \,\, C_n &=& \frac{1}{n!}\int {\rm Tr}(\frac{\mathcal{F}}{2\pi})^n \nn \\ &=& \frac{1}{2^n n!(2\pi)^n}\int d^{2n}k \epsilon^{\alpha_1\cdots\alpha_{2n}} {\rm Tr} \mathcal{F}_{\alpha_1\alpha_2}\cdots \mathcal{F}_{\alpha_{2n-1}\alpha_{2n}}
\label{chern-n} \eea which is a straightforward generalization of Eq.(\ref{chern}). Taking $n=1$ in Eq.(\ref{chern-n}), we have Eq.(\ref{chern}) with a slightly different notation, namely that the trace in Eq.(\ref{chern-n}) has been absorbed into the definition of $\mathcal{A}_i$ in Eq.(\ref{chern}).
The mathematical form in Equation.(\ref{chern-n}) is the same as that of the noninteracting Chern number\cite{thouless1982,nakahara1990,ryu2010}, but it is defined in terms of $h_t(\bk)$, thus it is applicable to interacting insulators and superconductors.

For topological insulators in class AIII in $2n+1$ dimensions, we have the chiral symmetry
\bea \{ h_t(\bk), \Gamma\} =0
\eea where $\Gamma$ satisfy $\Gamma^2=1$. It follows that $h_t$ can be written as \bea h_t(\bk) = \left( \begin{array}{cc}
 & \mathcal{Q}(\bk) \\
\mathcal{Q}^\dag(\bk) & \\
\end{array} \right)
\eea in the basis in which $\Gamma$ is diagonal. The topological invariant is given as a winding number \bea {\rm AIII}: \, W_{2n+1} &=& \frac{(-1)^n n!}{(2n+1)!}(\frac{i}{2\pi})^{n+1}\int {\rm Tr} (\mathcal{Q}^{-1}d\mathcal{Q})^{2n+1} \nn \\ &=& \frac{(-1)^n n!}{(2n+1)!}(\frac{i}{2\pi})^{n+1}\int d^{2n+1}k \epsilon^{\alpha_1\cdots\alpha_{2n+1}} \nn \\ &&\times  {\rm Tr} [( \mathcal{Q}^{-1}\partial_{\alpha_1}\mathcal{Q})\cdots ( \mathcal{Q}^{-1}\partial_{\alpha_{2n+1}}\mathcal{Q})] \label{winding} \eea
which takes the same form as the noninteracting winding number\cite{schnyder2008,ryu2010}.

We have listed formulas for the two complex classes, A and AIII. The eight real classes are similar. For topological insulators/superconductors in AI and AII classes in $4n$($n$ is an arbitrary positive integer) dimensions, and those in C and D classes in $4n-2$ dimensions, the topological invariants are Chern numbers with the same forms as given by Eq.(\ref{chern-n}).  For topological insulators/superconductors in CI and DIII classes in $4n-1$($n$ is an arbitrary positive integer) dimensions, and those in BDI and CII classes in $4n-3$ dimensions, the topological invariants are the winding numbers, whose forms are the same as Eq.(\ref{winding}). This list has exhausted integer topological invariants (Chern numbers and winding numbers).

The $Z_2$ topological invariants can be obtained from the dimensional reductions of integers invariants. This has been done for noninteracting insulators\cite{qi2008,ryu2010}, and can also be generalized to interacting insulators using $h_t$. For instance, the time reversal invariant topological insulators in AII class in 3D and 2D can be obtained from dimensional reduction of 4D topological insulator in the same class\cite{qi2008}, in other word, the $Z_2$ topological invariant in 3D and 2D (the quantum spin Hall insulator) can be obtained from the 4D Chern number.  The details of this dimensional reduction in the zero frequency Green's function approach can be found in Ref.\cite{wang2012a}. For the first descendents (i.e. time reversal invariant topological insulators in 3D), we can extend $h_t(\bk)$ to $h_t(\bk,u)$ by adding a Wess-Zumino-Witten parameter $u$ (similar to the procedure in Ref.\cite{wang2010b}). Now we have four variables $(k_1,k_2,k_3,u)$, thus we can define the second Chern number $C_2$, and identify the 3D $Z_2$ topological invariant as $C_2$ (mod $2$).   For the second descendents (i.e. time reversal invariant topological insulators in 2D), we can extend $h_t(\bk)$ to $h_t(\bk,u,v)$, and the rest procedure is the same. More generally, for interacting $Z_2$ topological insulators/superconductors in any spatial dimensions in which they do exist, we can implement a Wess-Zumino-Witten extension of $h_t$, so that the topological invariant of a topological insulator in lower dimension can be compactly defined in terms of quantities in higher dimensions, though the final result is independent (mod even integer) of this dimensional extension.
For the first descendents of Chern insulators, the result of dimensional reduction is exactly the Chern-Simons terms\cite{qi2008,ryu2010,nakahara1990}, therefore, there is no need to take the Wess-Zumino-Witten extension in calculation. For the descendents of ``winding number insulators'', we can take the Wess-Zumino-Witten terms as the definition of $Z_2$ topological invariants, though no simple form analogous to the Chern-Simons term can be found.

Since all these $Z_2$ topological invariants take the same forms as integer topological invariants,  we do not need to list all of them. Let us just take the DIII class as an example to illustrate how the integer topological invariants in $D$ dimensions can be regarded as $Z_2$ invariants in $D-1$ and $D-2$ dimensions.
For topological insulators/superconductors in DIII class in $4n-1$ dimensions, we have stated that the topological invariant is given by the winding number , namely that we just replace the dimension ``$2n+1$'' in Eq.(\ref{winding}) by ``$4n-1$''. Now we can write down the $Z_2$ topological invariant in $4n-2$ dimension as
\bea  {\rm DIII}: \, Z_2  &=&  \frac{  (2n-1)! (-1)^{n+1}}{(4n-1)!}(\frac{1}{2\pi})^{2n} \int {\rm Tr} (\mathcal{Q}^{-1}d\mathcal{Q})^{4n-1} \nn \\ &=&  \frac{  (2n-1)! (-1)^{n+1}}{(4n-1)!}(\frac{1}{2\pi})^{2n}  \int  du d^{4n-2}k \epsilon^{\alpha_1\cdots\alpha_{4n-1}} \nn \\ &&\times  {\rm Tr} [( \mathcal{Q}^{-1}\partial_{\alpha_1}\mathcal{Q})\cdots ( \mathcal{Q}^{-1}\partial_{\alpha_{4n-1}}\mathcal{Q})] \label{winding-z2} \eea where the variables are $k_1,\cdots,k_{4n-2}, u$. Here $u$ is a Wess-Zumino-Witten-extension parameter. More precisely, we have extended $\mathcal{Q}$ from the Brillouin zone $(k_1,\cdots,k_{4n-2})$ to $(k_1,\cdots,k_{4n-2},u)$($u\in[-1,1]$), with $(k_1,\cdots,k_{4n-2},0)$ identified as the original Brillouin zone $(k_1,\cdots,k_{4n-2})$[see Ref.\cite{wang2010b} for analogous treatment]. We can see that Eq.(\ref{winding-z2}) is exactly the winding number in $4n-1$ dimension if we regard $u$ as an additional momentum $k_{4n-1}$. The only subtlety is that Eq.(\ref{winding-z2}) is defined mod $2$, in other words, it is a $Z_2$ topological invariant. This (mod $2$) ambiguity is a well known fact of the Wess-Zumino-Witten terms\cite{witten1983}. For $4n-3$ dimensional topological insulators/superconductors in class DIII, the $Z_2$ topological invariant can be written down just like Eq.(\ref{winding-z2}), except that we have two parameters $u$ and $v$ in the Wess-Zumino-Witten extension.

All other $Z_2$ topological invariants can also be obtained from Eq.(\ref{chern-n}) and Eq.(\ref{winding}), with some momentum variables replaced by Wess-Zumino-Witten parameters. The resultant formulas are analogous to Eq.(\ref{winding-z2}). Because the translation from Eq.(\ref{chern-n}) and Eq.(\ref{winding}) to these $Z_2$ topological invariants is straightforward, we do not list all formulas here.

Equation (\ref{chern-n}), Equation (\ref{winding}), and their $Z_2$ descendents (through Wess-Zumino-Witten extensions) are topological invariants expressed in terms of $h_t$ for interacting systems.
These topological invariants do not involve frequency integral and are much easier to calculate than those with frequency integral.

Although the Green's function has rich behaviors of frequency dependence, it turns out that all information about topology is fully contained in its zero frequency value, which has been proven using the Lehmann spectral representation of Green's function and a smooth deformation\cite{wang2012a}. This mathematical fact established the zero frequency Green's function as a precise tool for topological invariants, since it implies that we cannot obtain more topological invariants even if we study Green's function in the entire frequency domain.  In the present paper, we confirm the zero frequency Green's function approach from the perspective of surface states, which is complementary to the calculations in the bulk\cite{wang2012a}.

\bibliography{Topo_Hamiltonian}

\end{document}